\documentstyle{aipproc2}
\begin{document}
\title{Observational Constraints on Inflation Models with Nonminimal Scalar
Field}

\author{Hyerim Noh$^*$ and Jai-chan Hwang$^{\dagger}$}
\address{$^*$Korea Astronomy Observatory,
Taejeon, Korea\\
$^{\dagger}$Department of Astronomy and Atmospheric Sciences,
Kyungpook National University, Taegu, Korea}

\maketitle

\begin{abstract}
The power spectra of the scalar- and tensor-type structures generated in an
inflation model based on nonminimally coupled scalar field are derived.
The contributions of these structures to the anisotropy of the cosmic
microwave background radiation  are derived, and are compared with
the four year COBE DMR data. 
The constraints on the ratio of the self-coupling and nonmimimal coupling
constants, the expansion rate during the inflation period, and the
relative amount of the tensor-type contribution to the quadrupole of the
CMBR temperature anisotropy are provided.
\end{abstract}

\section*{Introduction}

The inflation has been recognized as the strong mechanism which
can provide  the origin of the
large scale structure generation and the
evolution in the universe. Therefore it is possible to probe
the early universe by using the observation of the large scale structures.
Especially,  the observed anisotropies of the cosmic microwave background 
radiation
(CMBR) in the large angular scale are very important to constrain the inflation
models which are usually based on the model scalar fields or the
generalized version of the gravity theories. 
Since the first detection of the CMBR temperature anisotropies by COBE (1992),
several theories of structure formation have been investigated, and owing to
the advances in the observational techniques these theories have been
tested with increasing precision. A host of new data including the recent
results from Boomerang \cite{Bernardis}, MAXIMA-I and the near future
experiments like MAP (2001), Planck surveyor (2007) will map the CMBR
in more detail and consequently can provide stronger constraints
on the inflation models. 

Recently, the importance of the generalized gravity theories in the
early universe has increased, and these theories arise from the
attempts to quantize the gravity or as the low-energy  limit of the
unified theories. So, many inflation models have been investigated based on the
generalized gravity theories \cite{{Starobinsky},{Spokoiny}}. 

In this work, we will investigate the constraints on the inflation model
based on the nonminimall coupled scalar field with a self interaction.
The chaotic inflation model is usually based on the minimally coupled scalar
field with self coupling. However, this model has some fine tuning
problem with $\lambda$ which must be unreasonably small in order to
be consistent with the observed quadrupole anisotropy.
The new chaotic inflationary scenario was proposed based on the nonminimally
coupled scalar field with a strong coupling assumption 
\cite{Fakir-Unruh} in which the severe constraint on $\lambda$ can be
relaxed by introducing a large value of $\xi$.
Most of works on this model have been done either using the
scalar-type structures \cite{{SBB},{Fakir-Unruh},{Makino-Sasaki}}, 
or the tensor-type structures
\cite{Komatsu-Futamase}.
We use both the scalar- and the tensort-type structure, and
derive the  constraints on the coupling constants and the
expansion rate during the inflation era comparing with the
four-year COBE DMR data.

\section*{Non-minimally coupled Scalar Field}
 
The Lagrangian for the generalized gravity theory is given by
\begin{equation}
   {\cal L} = \sqrt{-g}
       \left[ {1 \over 2} f (\phi, R)
       - {1 \over 2} \omega (\phi) \phi^{;c} \phi_{,c}
       - V (\phi) \right].
\end{equation}
The non-minimally coupled scalar field is one case
with $f = (\kappa^{-2} - \xi \phi^2)R$,
thus $F \equiv f_{,R} = \kappa^{-2} - \xi \phi^2$, and $\omega = 1$ where
$\kappa^2 \equiv 8 \pi m_{pl}^{-2}$.
We consider a self-coupling $V = {1 \over 4} \lambda \phi^4$.

We assume the slow-rolls ($|\ddot \phi /\dot \phi| \ll H \equiv \dot a/a$
and $|\dot \phi/\phi| \ll H$),
the potential-dominance (${1 \over 2} (1 - 6 \xi) \dot \phi^2 \ll V$),
and the strong-coupling ($|\kappa^2 \xi \phi^2| \gg 1$) conditions, and
then obtain the following background solutions,
\cite{{Fakir-Unruh},{Makino-Sasaki},{Komatsu-Futamase}}:
\begin{equation}
   H = H_i + {\lambda \over 3 \kappa^2 \xi ( 1 - 6 \xi)}
       \left( t - t_i \right), \quad
       \phi = \sqrt{ - 12 {\xi \over \lambda} } H,
   \label{BG}
\end{equation}
where we consider $\xi < 0$ case \cite{Barvinsky}.
In the regime $H_i$ term {\it dominates} $H$
we have
a near exponentially expanding period $a \propto e^{H_i t}$ which can
provide a possible inflation scenario, \cite{Fakir-Unruh}.

Using our previous work in which we have investigated  the quantum generation
and the classical evolution  of the scalar- and tensor-type structures in
the generalized gravity theories,
we derive the general power spectra
based on vacuum expectation values in Eq. (16) of \cite{GGT-scalar}
and Eq. (32) of \cite{GGT-GW} in the large scale limit
\begin{equation}
     {\cal P}_{\hat \varphi_{\delta \phi}}^{1/2}
       = {H \over |\dot \phi|} {\cal P}_{\delta \hat \phi_\varphi}^{1/2}
       = {H^2 \over 2 \pi |\dot \phi|} {1 \over \sqrt{1 - 6 \xi} }
       \big| c_2 (k) - c_1 (k) \big|,
   \label{P-scalar-general}
\end{equation}
\begin{equation}
     {\cal P}_{\hat C_{\alpha\beta}}^{1/2}
       = {\kappa H \over \sqrt{2} \pi} {1 \over \sqrt{1 - \kappa^2 \xi \phi^2} }
       \sqrt{ {1 \over 2} \sum_\ell
       \big| c_{\ell 2} (k) - c_{\ell 1} (k) \big|^2 },
   \label{P-GW-general}
\end{equation}
where $\ell$ represents the two polarization states of the gravitational wave.
We find these results reproduce the minimally coupled scalar field case
in the limit of  $\xi = 0$,
\cite{Stewart-Lyth}.
$c_i(k)$ and $c_{\ell i}(k)$ are constrained by the quantization
conditions: $|c_2|^2 - |c_1|^2 = 1$, and $|c_{\ell 2}|^2 - |c_{\ell 1}|^2 = 1$.
We see that the power spectra  generally depend on the scale $k$
through the vacuum choices which fix $c_i$ and $c_{\ell i}$.
If we choose the simplest vacuum states $c_2 = 1$ and $c_{\ell 2} = 1$
the power spectra become independent of $k$, thus are scale invariant.

Eq. (\ref{BG}), Eqs. (\ref{P-scalar-general}), and (\ref{P-GW-general}) give:
\begin{equation}
     {\cal P}_{\hat \varphi_{\delta \phi}}^{1/2}
       = \left( {H_i \over m_{pl}} \right)^2
       \sqrt{ - { 12 \xi ( 1 - 6 \xi) \over \lambda} }
       \big| c_2 - c_1 \big|,
   \label{P-scalar-NM-quantum} 
\end{equation}
\begin{equation}
     {\cal P}_{\hat C_{\alpha\beta}}^{1/2}
       = {1 \over 2 \pi} \sqrt{\lambda \over 6 \xi^2}
       \sqrt{ {1 \over 2} \sum_\ell \big| c_{\ell 2} - c_{\ell 1} \big|^2 }.
   \label{P-GW-NM-quantum}
\end{equation}
Using our previous results that  ${\varphi_{\delta\phi}}$ and 
$C_{\alpha\beta}$ are conserved independently of changing gravity theory,
changing potential, and changing equation of state in the large scale,
we can identify the power spectra based on the vacuum expectation values
during the inflation with the classical ones based on the spatial averaging.
So, the Eq. (\ref{P-scalar-NM-quantum}), (\ref{P-GW-NM-quantum}) are
valid even in the matter dominated era in the large scale. 
Consequently the classical power spectra during the matter dominated era are 
\begin{equation}
     {\cal P}_{\varphi_{\delta \phi}}^{1/2}
       = \left( {H_i \over m_{pl}} \right)^2
       \sqrt{ - { 12 \xi ( 1 - 6 \xi) \over \lambda} },
  \label{P-scalar}
\end{equation}
\begin{equation}
     {\cal P}_{C_{\alpha\beta}}^{1/2}
       = {1 \over 2 \pi} \sqrt{\lambda \over 6 \xi^2}.
   \label{P-GW}
\end{equation}

Comparing these power spectra with the observed data of the
quadrupole anisotropy  in the CMBR temperature we can derive important
constraints on the inflation model parameters.
For example, the observed values of 
$\langle a_2^2 \rangle$ are related with the power spectra
\cite{{GGT-scalar},{GGT-GW}}, and consequently constrain 
$H_i$ and $\lambda/\xi^2$.

{}For the scale independent Zel'dovich spectra 
in Eqs. (\ref{P-scalar},\ref{P-GW})
the quadrupole anisotropy is given by
\begin{equation}
   \langle a_2^2 \rangle
   = \langle a_2^2 \rangle_S + \langle a_2^2 \rangle_T
   \nonumber \\
   = {\pi \over 75} {\cal P}_{\varphi_{\delta \phi}}
       + 7.74 {1 \over 5} {3 \over 32} {\cal P}_{C_{\alpha\beta}}.
   \label{a_2}
\end{equation}
The four-year {\it COBE}-DMR data give \cite{COBE}:
\begin{equation}
     Q_{\rm rms-PS} = 18 \pm 1.6 \mu K, \quad
       T_0 = 2.725 \pm 0.020 K,
   \quad
     \langle a_2^2 \rangle
       = {4 \pi \over 5} \left( {Q_{\rm rms-PS} \over T_0} \right)^2
       \simeq 1.1 \times 10^{-10}.
   \label{a_2-value}
\end{equation}

We consider a case
with $|\xi| \gg 1$.
The ratio of two types of the structures is given by
\begin{equation}
  r_2 \equiv \langle a_2^2 \rangle_T / \langle a_2^2 \rangle_S
    =3.46{\cal P_{C_{\alpha\beta}}}/{\cal P_{\varphi_{\delta \phi }}}.
\end{equation}

The additional constraints can be obtained by using the
e-folding number $N_k$  for the successful inflation.
$N_k$ is the number of e-folds since the scale $k$ exists the
horizon and reachs the large scale  limit before the end of the latest
inflation and is defined by
\begin{equation}
   N \sim \int_{t_k}^{t_e} H dt
             \sim  -{H^2 \over 2\dot H} (t_k),
   \label{N-def}
\end{equation}
where $t_e$ is the ending epoch of the latest inflation, and
$t_k$ is the epoch when the perturbation with the scale $k$ exits the
horizon and reaches the large scale. 

Using $N_k \sim 60$ gives following constraints:
\begin{equation}
   r_2  = 3.46 {3\over N_k^2} \sim 0.0029,
   \quad
   {\lambda \over \xi^2 } \sim 5.2 \times 10^{-10},
   \quad 
   {H(t_k) \over m_{pl}} \sim \sqrt {75 \langle a_2\rangle^2 /N_k}
   \sim 1.2 \times 10^{-5}.
\end{equation}
These results show the strong constraint.
Also, we derive the spectral indices of the scalar and tensor type
structures
\begin{equation}
    n_S -1 = -2/N_k \sim -0.033, \quad
    n_T \sim 0.
\end{equation}
Therefore our results give the nearly scale independent spectra which are
consistent with current observations.

\section*{Discussion}

We derive several constraints including the ratio of the parameters
and the expansion rates during the inflation era.
Futhermore, using the condition of the successful inflation with
enough e-folds we obtain the strong constraints on the gravitational
wave contribution to the quadrupole temperature anisotropy of the CMBR.
The gravitational wave contributon turns out to be negligible 
compared with the scalar-type contribution.
Therefore any future observational data which show the excessive amount of
the gravitational wave contribution may exclude the possibility
of this model.

Our results also apply to the case of induced gravity  in which
$f = \epsilon \phi^2 R$, $\omega = 1$, and
$V = {1 \over 4} \lambda (\phi^2 - v^2 )^2$.
Assuming that $\phi^2 \gg v^2$, our
nonminimally-coupled scalar field
with strong coupling and $V = {1 \over 4} \lambda \phi^4$ becomes
exactly the induced gravity case.
By replacing $\xi \rightarrow - \epsilon$ and $\xi \kappa^2 \rightarrow - v^2$,
our analyses and results apply to the inflation based on
induced gravity.


\begin{references}
\bibitem{Bernardis}
         de Bernardis, P., {\it et al.}, {\it Nature}\ {\bf 404}, 955 (2000). 
\bibitem{Starobinsky}
         Starobinsky, A. A., {\it Phys. Lett. B}\ {\bf 91}, 99 (1980).
\bibitem{Spokoiny}
         Spokoiny, B. L., {\it Phys. Lett. B}\ {\bf 147}, 39 (1984).
\bibitem{Fakir-Unruh}
         Fakir, R., Unruh, W. G., {\it  Phys. Rev. D}\  {\bf 41}, 1783 (1990);
         {\it ibid.} {\bf 41}, 1792 (1990);
         Fakir, R., Habib, S., and Unruh, W. {\it  Astrophys. J.}\
         {\bf 394}, 396 (1992).
\bibitem{SBB}
         Salopek, D. S., Bond, J. R., and Bardeen, J. M., 
         {\it  Phys. Rev. D}\ {\bf 40}, 1753 (1989).
\bibitem{Makino-Sasaki}
         Makino, N., and Sasaki, M., {\it Prog. Theor. Phys.}\ 
         {\bf 86}, 103 (1991).
\bibitem{Komatsu-Futamase}
         Komatsu, E., and Futamase, T., {\it Phys. Rev. D}\ 
         {\bf 58}, 023004 (1998).
\bibitem{Barvinsky}
         Barvinsky, A. O., Kamenshchik, A. Yu., and Mishakov,
         {\it Nucl. Phys. B}\ {\bf 491}, 387 (1997).
\bibitem{GGT-scalar}
         Hwang, J., and Noh, H., {\it Class. Quant. Grav.}\ {\bf 15},
         1387 (1998).
\bibitem{GGT-GW}
         Hwang, J., {\it Class. Quant. Grav.}\ {\bf 15}, 1401 (1998).
\bibitem{Stewart-Lyth}
         Stewart, E. D., and Lyth, D. H., {\it Phys. Lett. B}\ 
         {\bf 302}, 171 (1993).
\bibitem{COBE}
         Bennett, C. L., {\it Astrophys. J.}\ {\bf 464}, L1 (1996);
         G\'orski, K. M., {\it et. al.}, {\it ibid.}\ {\bf 464}, L11 (1996);
         G\'orski, K. M., {\it et. al.}, {\it Astrophys. J. Suppl.}\
         {\bf 114}, 1 (1998).
\end{references}
\end{document}